\documentclass[journal=amchemso,manuscript=article,]{achemso}

\usepackage[english]{babel}

\usepackage{amsmath}
\usepackage{graphicx}
\usepackage[colorlinks=true, allcolors=blue]{hyperref}
\usepackage{float} 

\usepackage{booktabs}

\usepackage{chngcntr}
\usepackage{multirow}

\SectionNumbersOn

\title{Black-Box Uncertainty Estimation for Deep Learning Models in Atomistic Simulations}

\author{Idan Fonea}
\affiliation{Department of Physical Electronics, School of ECE, Tel Aviv University, Israel}
\email{idanfonea@gmail.com}
\author{Amir Peles}
\affiliation{Department of Physical Electronics, School of ECE, Tel Aviv University, Israel}
\author{Sivan Niv}
\affiliation{Department of Physical Electronics, School of ECE, Tel Aviv University, Israel}
\author{Goren Gordon}
\affiliation{School of Industrial Engineering and Intelligent Systems, Tel Aviv University, Israel}
\alsoaffiliation{Luddy School of Informatics, Computing, and Engineering, Indiana University Bloomington, Bloomington, IN, USA}

\author{Amir Natan}
\affiliation{Department of Physical Electronics, School of ECE, Tel Aviv University, Israel}
\alsoaffiliation{The Sackler Center for Computational Molecular and Materials Science, Tel Aviv University, Israel}
\email{amirnatan@post.tau.ac.il}

\begin{document}
\maketitle

\begin{abstract}
We analyze an ensemble-based approach for uncertainty quantification (UQ) in atomistic neural networks. This method generates an epistemic uncertainty signal without requiring changes to the underlying multi-headed regression neural network architecture, making it suitable for sealed or black-box models. We apply this method to molecular systems, specifically sodium (Na) and aluminum (Al), under various temperature conditions. By scaling the uncertainty signal, we account for heteroscedasticity in the data. We demonstrate the robustness of the scaled UQ signal for detecting out-of-distribution (OOD) behavior in several scenarios. This UQ signal also correlates with model convergence during training, providing an additional tool for optimizing the training process.

\end{abstract}

\section{Introduction}

In recent years, machine learning (ML), and specifically neural networks (NN), has offered promising approaches for enhancing and potentially replacing quantum calculations, such as Density Functional Theory (DFT) in molecular simulations. Those models for the system's energy and the atomic forces are orders of magnitude faster to calculate but yield a highly accurate approximation. These NN models are trained on datasets of atomistic configurations to predict energy and forces that are satisfactorily close to DFT results. A common NN architecture for the calculation of atomic forces was suggested by Bheler and Parrinello~\cite{behler2007generalized}, in their model the total energy of the system is first calculated and then the forces are derived from the energy gradients. Since then, numerous variations of NN architectures for the prediction of the system total energy and atomic forces were used in molecular simulations.~\cite{smith2017ani,von2018quantum,gassner1998representation,manzhos2006nested,lorenz2004representing, rupp2012fast, montavon2013machine, hansen2015machine,behler2011atom}. It is also possible to build NN models that calculate the forces directly without calculating the energy of the system~\cite{kuritz2018size,sivanniv2020}.

A known limitation of most NN models for quantum calculations, whether energy-based, or direct forces-based, is that they tend to suffer from generalization errors. This means that while NN models can accurately predict properties for a dataset that is similar to the training dataset, they may produce large errors when the atomistic configuration data distribution deviates significantly from it. An extremely important example is the simulations of surface reactions and surface catalysis where the end geometry might deviate significantly from the initial state of the simulation. This poses a need for the prediction of not only the required physical properties and quantity of interest but also a measure of the model's reliability. This measure is typically referred to as predictive uncertainty, which quantifies the uncertainty of an ML model's prediction. It usually denotes a proxy signal that produces an estimate or some correlated information about the error in the model's prediction.

The statistical behavior of the data, and products of it, such as the prediction, the error, or the uncertainty estimation, is commonly divided into two types, homoscedastic and heteroscedastic. Typically, homoscedastic uncertainty statistical properties do not vary as a function of the input space, whereas heteroscedastic uncertainty properties do.

 Predictive uncertainty can also be further divided into two types: aleatoric and epistemic~\cite{hullermeier2021aleatoric}. Aleatoric uncertainty is often attributed to randomness in the data and is often considered to be a representation of the intrinsic errors or noise in the dataset or the inadequacy of the model architecture to interpolate the data points (which is impossible to overcome without changing the model's architecture)~\cite{hullermeier2021aleatoric}.

Several methods have been proposed to estimate aleatoric uncertainty, including predicting, in addition to the target, the parameters of the prediction distribution (such as the standard deviation) or use of a modified loss, such as ping-pong loss, to infer upper or lower bounds for the error~\cite{tagasovska2019single}. 

In contrast to aleatoric, epistemic uncertainty is often attributed to the trained model alignment to the presented data; It can be reduced with a change in the training data. Various methods have been proposed to estimate epistemic uncertainty, including the use of deep ensembles~\cite{lakshminarayanan2017simple} and MC-dropout~\cite{gal2016dropout}. These approaches were implemented in the field of ML-based MD by Musil et al.~\cite{musil2019fast} and by Wen and Tadmor~\cite{wen2020uncertainty}, the performance of those different methods was analyzed by Hirschfeld et al.~\cite{hirschfeld2020uncertainty}. Another approach that was presented and analyzed by Janet et al.~\cite{janet2019quantitative} is to use distance metrics from the training data representation or latent variables.

Most approaches for uncertainty quantification (UQ) require the modification of an existing NN model, by using a dropout layer, by modifying the model to predict additionally the standard deviation, or by tapping into the network architecture to extract the latent parameters.  While such methods can be highly effective, they might fail in important cases~\cite{seitzer2022pitfalls}. In addition, they are not applicable to a sealed model in a model-agnostic manner. Hence, they cannot be applied to an existing model that cannot be changed. Furthermore, in some cases, there is a hidden assumption of a given statistical behavior of the input data, which may lead to a model over-confidence or a model under-confidence of the predicted variable, when the statistical behavior of the data strays from the training set. Although the ensemble method is more computationally demanding in comparison to MC dropout, it was found by some studies to be more accurate~\cite{janet2019quantitative,ovadia2019can,dietterich2000ensemble,lakshminarayanan2017simple}. In addition, the ensemble method is trivially parallelized.
It is important to note, however, that some errors, such as a constant bias, can be missed by all ensemble-based methods~\cite{carrete2023deep}.

For an efficient learning, samples with a high uncertainty can be selected for additional calculation with DFT and re-training, as suggested by Smith et al.~\cite{smith2018less}, such samples, with a high UQ, can be viewed as samples that add more information to the model, hence improve it.

In our study, we produce a UQ signal for the epistemic uncertainty, to determine whether the trained NN model is reliable, or whether we need to add additional labeled data or modify the training dataset and retrain the model. We preferred to use the ensemble method over the MC-Dropout and other methods for the production of the UQ signal. This is mostly because our chosen ensemble method allows us to use the NN model as it is without the need to alter the model to produce the UQ signal. Hence, this method can be applied to any single-headed or multi-headed regression model.

We create a UQ signal from an ensemble of an existing model, which is measured based only on the ensemble predictions disagreement~\cite{smith2018less,musil2019fast,lakshminarayanan2017simple}.
Specifically, the estimation is done by using an ensemble of predictors taken as a black box, without changing their model architecture or training loss, but with training each expert using a sub-sample of the total train data (also referred to as bootstrapping). For the specific tested NN architecture for the forces we used previous models that were developed in our group~\cite{kuritz2018size,sivanniv2020}.

After implementing a UQ signal for an existing architecture and dataset, we empirically determine a threshold for the UQ signal, above which, the data is assumed to be out of distribution (OOD). In such a case, the prediction is assumed to be fault-prone, and additional labeled data should be gathered using new DFT simulations.

We discuss the performance of this UQ signal and model error for several cases. We first analyze a model trained on data that was generated for bulk sodium and aluminum with MD at one temperature for the prediction of forces on data of MD at other temperatures. We then analyze the behavior of those parameters also for the case of a model trained on bulk aluminum and then tested with new data containing surfaces. Finally, We  analyze the performance of the model error and UQ signal as a function of the number of data samples.

We show that our UQ signal has the ability to separate between data populations, furthermore, we show that due to the heteroscedastic nature of the data, some scaling of the UQ signal should be considered. 

Since a heteroscedastic data behavior detection by the UQ signal can be encountered in various fields and various NN models, this analysis can be of quite general scope and could be useful for other problems where a black box model's prediction validity estimation is desired.

\section{Methods}
\subsection{Datasets}

The dataset was produced with {\it ab initio} MD (AIMD) quantum simulations that were implemented with the VASP DFT package~\cite{kresse1993ab,kresse1996efficient}. To create the datasets, for sodium (Na) we used a starting point a body-centered (BCC) super-cell of 3x3x3, containing 27 atoms with periodic boundary conditions, with the ground state lattice constant.  For aluminum (Al) we used a face-centered cubic (FCC) super-cell with 27 atoms. The PBE~\cite{PBE_PhysRevLett.77.3865}  functional was used throughout without spin polarization, the VASP PAW pseudopotentials~\cite{blochl_PhysRevB.50.17953,kresse_PhysRevB.59.1758} were used, with a $K$-point grid of $5\times5\times5$ and cutoffs of 260 eV for sodium and 360 eV for aluminum.  NVT ensemble simulations were performed using the Nos\'e--Hoover thermostat~\cite{nose1984unified,hoover1985canonical} with temperatures ranging from 100$\,\mathrm{K}$ to 4000$\,\mathrm{K}$ (each temperature was simulated separately into a data set). The MD simulations were calculated with a 2fs time step, and the number of data points is roughly 1000 for each temperature, where each data point is a cell snapshot in time. 

\subsection{Network architecture}

\begin{figure}
	\centering
	\includegraphics[width=0.75\textwidth]{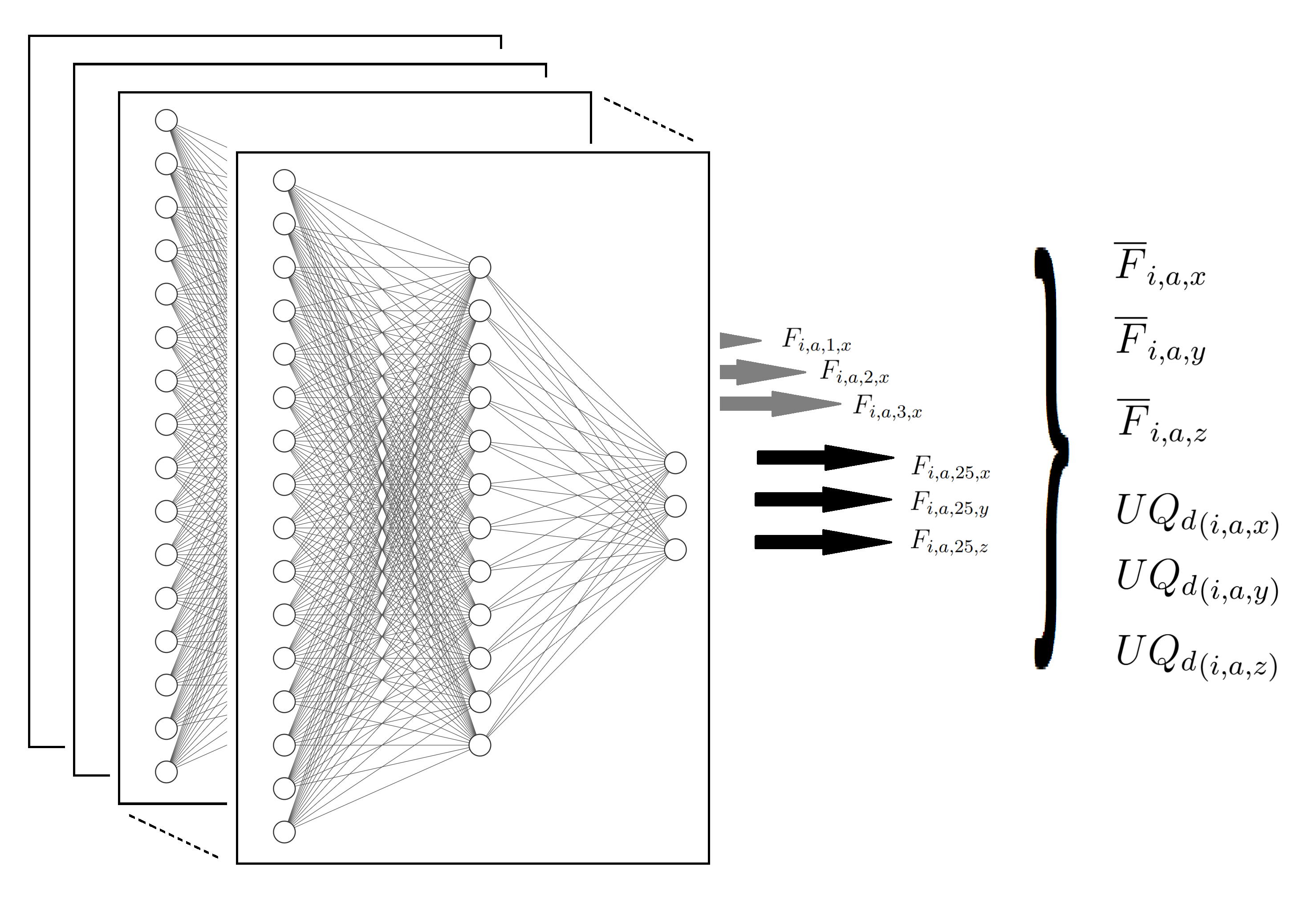}
	\caption{
		Ensemble of $25$ neural networks for force prediction and uncertainty quantification.
		Each network, trained on a sampled dataset, predicts the Cartesian force components 
		$\vec{F}_{i,a,k}$  for atom $a$ on data sample $i$. 
		Ensemble averaging yields the mean forces $\vec{\overline{F}}_{i,a}$ 
		and standard deviation across models $\vec{UQ_d}(i,a)$.
	}\label{fig:figure1_dnn_diag}
	
\end{figure}

We tested our UQ signal using an ensemble of a deep learning model that directly predicted the forces~\cite{kuritz2018size,sivanniv2020}. For each atom, the model's input is a list of the neighbor atom distance vectors. This set of vectors is then transformed into a feature set in a fashion inspired by Zhang et al.~\cite{zhang2018end} This feature set is then fed to a dense neural network which produces the forces directly. The details of the model architecture and parameters are provided in the supporting information (SI). 

We used an ensemble of 25 identical architecture models (experts), as shown in Figure \ref{fig:figure1_dnn_diag}, each trained on a random sampling of 80\% (bootstrap) of the train set, giving 64\% data overlap between each expert in the ensemble.

{This bootstrap approach (also referred to as
	bootstrap aggregation~\cite{carrete2023deep}) was chosen over a simpler committee setup
	(i.e., training on identical data with different random initializations~\cite{carrete2023deep})
	to better probe the model's sensitivity to data representation. 
	Our assumption was that this sub-sampling of the training data set would more strongly affect the ensemble disagreement on test samples that are under-represented in the training dataset.  }

The uncertainty is then obtained by measuring the spread of the force predictions of the ensemble for a single atom in the cell. Additional scaling and smoothing operations are performed on the UQ signal as detailed in the sections below. This bootstrap approach for building the experts does not rely on the particular network architecture and therefore can be applied to other models.

\subsection{Direct UQ signal formulation}

We define the ensemble prediction, ${\vec{\overline{F}}_{i,a}}$, as the average prediction of all the experts at data sample ${i}$ on atom ${a}$. We use this ensemble prediction as the model prediction, and we calculate both the error and uncertainty relative to it. ${\vec{\overline{F}}_{i,a}}$ can be written as: 

\begin{equation}
\vec{\overline{F}}_{i,a} = \frac{1}{M} \sum_{m=1}^{M} \vec{F}_{i,a,m} \label{eq1:avgForce}
\end{equation}

Where ${M}$ is the number of experts in the ensemble ($M=25$ in our case), ${m}$ is the expert id, and ${\vec{F}_{i,a,m}}$ is the 3D force prediction of expert $m$ for data sample  ${i}$ and atom ${a}$.

We define the absolute value error (AE) in the ensemble force prediction as:

\begin{equation}
AE_{i,a}=   \| \vec{\overline{F}}_{i,a} - \vec{F}^{DFT}_{i,a} \| _{{_{_{1}}}} \label{eq:AE_defintion_mb2}
\end{equation}
Where ${\vec{F}^{DFT}_{i,a}}$ is the DFT calculated force label at sample $i$ on atom ${a}$ and the $\lVert\cdot\rVert_1$  norm is defined as 
$\| {\vec{F}}\|_1 = \lvert {F_x}\rvert + \lvert {F_y}\rvert + \lvert {F_z}\rvert$

We define the mean absolute error (MAE) for a dataset by averaging $AE_{i,a}$ over all the atoms in the cell and over all data samples~\cite{kuritz2018size}:

\begin{equation}
{MAE} =  \frac{1}{3S}\frac{1}{A}\sum_{i=1}^{S}\sum_{a=1}^{A}\| \vec{\overline{F}}_{i,a} - \vec{F}^{DFT}_{i,a} \| _{{_{_{1}}}} \label{eq:mae_pb3}
\end{equation}

Where ${S}$ is the number of samples in the measured data and ${A}$ is the number of atoms in the unit cell.

A common approach for the definition of the UQ signal is to take the standard deviation of the experts predictions. Another common method is to take the "half-spread" signal, defined as the half of difference between the 90th percentile and the 10th percentile of the predictions as proposed by Peterson et al.~\cite{peterson2017addressing} for a similar use case. We tried both approaches and found similar performance, we hence decided to focus on the standard deviation and use it for the UQ signal throughout this work. We define the (direct) vector UQ signal $\vec{UQ_d}_{(i,a)}$ for data sample $i$ on atom $a$ by\footnote{Note that in Eq. \ref{eq:UQd_signal_mb4} we divided by $M$ and not by $M-1$. The only effect on the results is a scaling by the constant $(M-1)/M$.}

\begin{equation}
UQ_{d_{(i,a,\beta)}}^2 = \frac{1}{M} \sum_{m=1}^{M} ({\overline{F}}_{i,a,\beta} - {F}_{i,a,m,\beta})^2 \label{eq:UQd_signal_mb4}
\end{equation}

where $\beta=x,y,z$ are the components of the Cartesian axes. Note that the definition of Eq. \ref{eq:UQd_signal_mb4} is purely local in time and in space. As before, $M$ is the number of experts.
{We also note that this definition of $UQ_d$, which being a measure of the variance of the ensemble,
is mostly considered sensitive to epistemic uncertainty.  As discussed in the introduction, this metric will not detect a
uniform model bias. 
}

\subsection{Relative AE and relative Uncertainty}

One way of evaluating the UQ signal is to compare it directly to the AE,  searching for mutual relationships and correlations. Another way is to convert the regression task to a classification problem, in which the data is labeled by whether it is in or out of distribution, based on the model train data.

As later described in the results section, we found that although the UQ signal (Eq. \ref{eq:UQd_signal_mb4}) correlates well with the AE (Eq. \ref{eq:AE_defintion_mb2}), it fails dramatically to detect out of distribution data, probably because both the AE and the UQ signal seems to correlate with the magnitude of predicted force.

We therefore searched for a relative or normalized definition for both the error and UQ signal. As they both scale with the force magnitude, a "natural" choice would be to scale both quantities by a factor in the scale of the force magnitude.

We hence define the relative AE, ${AE_r}_{i,a}$, by: 

\begin{equation}
{AE_{r_{i,a}}}=\frac{\| \vec{\overline{F}}_{i,a} - \vec{F}^{DFT}_{i,a} \|_{{_{_{1}}}}}{\| \vec{\overline{F}}_{i,a} \|_{{_{_{1}}}} + \delta} \label{eq:rel_AE_mb5}
\end{equation}

Here, $\delta$ is chosen to be a small enough value (we used $10^{-2} eV/\AA$) to avoid a division by zero. We also define the Mean Relative AE ($MAE_r$) by averaging the relative AE over all data samples and atoms: 

\begin{equation}
{
{MAE_r} = \frac{1}{3S} \frac{1}{A} \sum_{i=1}^{S}\sum_{a=1}^{A}  \frac{\| \vec{\overline{F}}_{i,a} - \vec{F}^{DFT}_{i,a} \|_{{_{_{1}}}}}{\| \vec{\overline{F}}_{i,a} \|_{{_{_{1}}}} + \delta }\label{eq5:relMAE_mb6}
}
\end{equation}

Where, as before, S is the number of samples, and $\vec{F}^{DFT}_{i, a}$ are the forces generated by the DFT calculation, which serves as ground truth. This metric is similar in nature to MAPE~\cite{de2016mean} { but we divide by the predicted value instead of the true value to be consistent with parameters that can be calculated during the inference time.}

 In a similar manner, we define the relative UQ (${UQ_r}$) signal by:

\begin{equation}
{\vec{UQ_r}_{(i,a)}} = {\frac{ \vec{UQ_d}_{(i,a)} }{ \overline{NF}_i + \delta} } \label{eq:U_calbirated_mb7}
\end{equation}

The $UQ_r$ signal, defined by Eq. \ref{eq:U_calbirated_mb7}, should have the force magnitude in the denominator, which can make it noisy if performed per atom per sample. To obtain a less noisy $UQ_r$ signal, we average the norm of the predicted forces over the unit cell and apply an additional sliding time-averaging window of width $W_n$, as described in Eqs.~\ref{eq:F_norm_mb8a} and~\ref{eq:F_norm_mb8}. The resulting average, $\overline{NF}_i$, is then inserted in denominator of Eq. \ref{eq:U_calbirated_mb7}, to receive the relative uncertainty. This unit-less ${UQ_r}$ signal is similar in spirit to the coefficient of variation (COV) of the ensemble, but with a smoothed  average of the Force norm, $\overline{NF}_i$, and the addition of ${\delta}$ in the denominator.

\begin{equation}
 \overline{NF}_{i,a} = \frac{1}{3W_{n}} \sum_{s=i-W_{n}+1}^{i} \|\vec{\overline{F}}_{s,a}  \|_{{_{_{1}}}}
\label{eq:F_norm_mb8a}
\end{equation}

\begin{equation}
 \overline{NF}_i = \frac{1}{A} \sum_{a=1}^{A} \overline{NF}_{i,a} 
\label{eq:F_norm_mb8}
\end{equation}

where $A$ is the number of atoms in the unit cell and ${a}$ represents an index of a specific atom, and $W_n$ is a sliding window for averaging over samples. In most cases we will use Eq. \ref{eq:U_calbirated_mb7} with $\overline{NF}_i$ in the denominator as defined in Eq. \ref{eq:F_norm_mb8}, in such cases we average the denominator over all the unit cell to reduce noise. In case we want to isolate the measurement to a specific atom, we may use instead the expression for ${\overline{NF}}_{i,a}$ as defined in Eq. \ref{eq:F_norm_mb8a}, and compensate for the decrease in the amount of averaged samples with a larger smoothing window ($W_n$). Finally, the $UQ_r$ signal, defined by Eq. \ref{eq:U_calbirated_mb7} is a vector quantity by construction, in most cases, when directional information is not important, we consider its norm.

We also further smoothed the whole ${UQ_r}$ signal by applying a sliding window of consecutive samples, resulting in a generalization of Eq. \ref{eq:U_calbirated_mb7} to Eq. \ref{eq:U_smoothed_calbirated_mb9}.

\begin{equation}
\vec{UQ}_{r(i,a)} = \frac{1}{W_{u}} \sum_{s=i-W_{u}+1}^{i}\frac{ \vec{UQ_d}_{s,a} }{ \overline{NF}_s + \delta}  \label{eq:U_smoothed_calbirated_mb9}
\end{equation}

In what follows we use Eq.\ref{eq:U_smoothed_calbirated_mb9} for the $UQ_r$ signal.
For bulk systems we discovered that the smoothing over cell atoms in Eq. \ref{eq:F_norm_mb8a} and Eq. \ref{eq:U_smoothed_calbirated_mb9}, choosing $W_n=W_u=1$, is enough.
Note that while the ${\vec{UQ_r}_{(i,a)}}$ definition of Eq. \ref{eq:U_calbirated_mb7} is purely local atom-wise, the ${\vec{UQ_r}_{(i,a)}}$ definition of Eq. \ref{eq:U_smoothed_calbirated_mb9}, while still local in the nominator, has an option in the denominator for averaging of the force magnitude over all the atoms in the cell. Depending on the use case, the $UQ_r$ signal, defined by Eq.  \ref{eq:U_smoothed_calbirated_mb9}, can be further averaged to get a $UQ_r$ estimation for the whole unit cell.

\subsection{UQ Signal Interpretation}

To use the uncertainty signal for decisions, we treat the observed uncertainty as a probability function distribution, similarly in spirit to conformal prediction~\cite{tibshirani2019conformal}. 
We take the UQ signal as the non-conformity score , obtained from predictions on test data of size $S_{test}$ with the same distribution of the training set (exchangeable), to derive a threshold ($TH$) that corresponds to the top $\alpha$ measured likelihood, where we took $\alpha = 2.5\%$ .

The Threshold is defined by:

\begin{equation}
TH(\alpha) = \arg\min_{TH}(|\frac{1}{S_{test}}\sum_{i=0}^{S_{test}-1}H({{UQ_r}_{(i,a,\beta)}},TH) - \alpha|)
\label{eq:Select_TH_mb10}
\end{equation}

Where $UQ_{r(i,a,\beta)}$ is the L1 norm of the $\vec{UQ}_r$ signal, and $H$ is the Heaviside function, and $i$ is the sample number, $\beta \in \{x,y,z\}$ is the Cartesian direction, $a$ is the atom index, and $TH$ is the threshold. Note that here the ${{UQ_r}_{(i,a,\beta)}}$ signal is taken as a local signal but can be smoothed and averaged, depending on the use case.

\begin{equation}
H({UQ_r}_{(i,a,\beta)}, TH) = 
\begin{cases} 
1, & \text{if } {{{UQ_r}_{(i,a,\beta)}}  \geq TH } \\
0, & \text{otherwise}
\end{cases}
\label{eq11_3:heavyside}
\end{equation}

Finally, we define the likelihood, P, of a test dataset of size ${S_{test}}$ being OOD of the train data by:

\begin{equation}
P_{OOD} =1-  \frac{1}{S_{test}}\sum_{i,a,d}^{S_{test}}H({UQ_r}_{(i,a,\beta)},TH)\label{eq:P_ood_mb12}
\end{equation}

\section{Results and Discussion}

In this section we show our analysis of both the $UQ_d$ and $UQ_r$ signals for several cases. We first discuss the performance of relative and direct UQ signals for the case of bulk sodium at several temperatures,  we then  demonstrate the same approach for bulk aluminum and for aluminum surfaces, finally we also check the behavior of the AE and UQ signal as a function of the training database size.

\subsection{Direct UQ signal temperature effects}

We initially tested the direct UQ signal, denoted by $\vec{UQ_d}_{(i,a)}$ in  Eq. \ref{eq:UQd_signal_mb4}, on models that were trained on datasets of bulk sodium (Na) at 300$\,\mathrm{K}$ (denoted by Na300$\,\mathrm{K}$) and at 2000$\,\mathrm{K}$ (denoted by Na2000$\,\mathrm{K}$). 

\begin{figure}[H]
    \centering
    \includegraphics[width=1.0\textwidth]{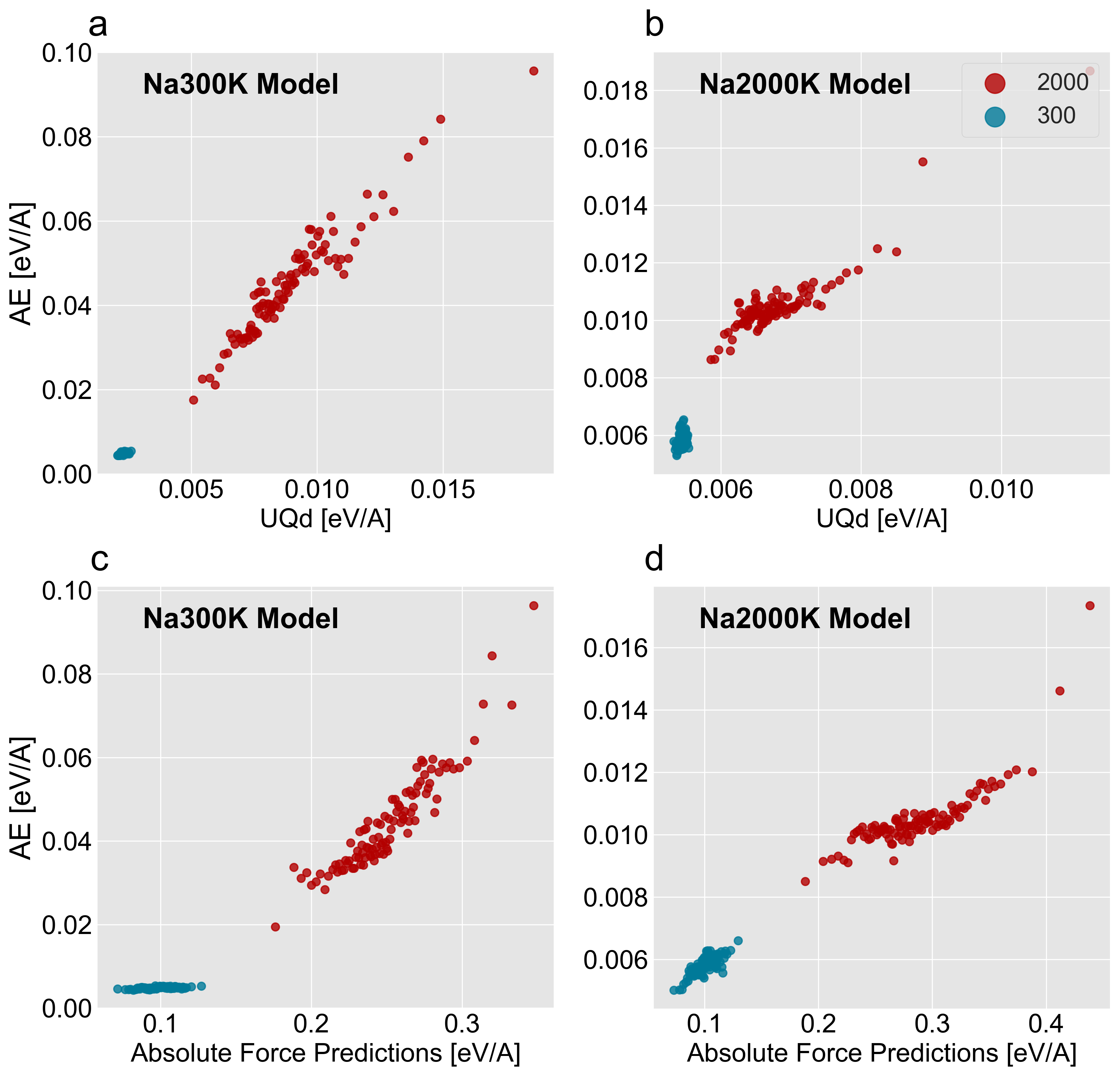}
    \caption{Panels (a, b) show a scatter plot of the AE (Eq.~\ref{eq:AE_defintion_mb2}) vs. the $UQ_d$ signal (Eq.~\ref{eq:UQd_signal_mb4}) produced for the 300$\,\mathrm{K}$ (a) and the 2000$\,\mathrm{K}$ (b) Na trained models, tested on 2000$\,\mathrm{K}$(red) and 300$\,\mathrm{K}$(blue) data.  Panels (c,d) show a scatter plot of the AE (Eq.~\ref{eq:AE_defintion_mb2}) vs the average absolute predicted force (Eq.~\ref{eq:F_norm_mb8}) is shown for models trained on 300$\,\mathrm{K}$ (c) and 2000$\,\mathrm{K}$ (d) of Na. Red is the result of a Na2000$\,\mathrm{K}$ test set and Blue is Na300$\,\mathrm{K}$ test set. In all graphs, the data is grouped into 100 bins according to the x-axis, and the AE is averaged in each bin.}
    \label{fig:scatter_U_and_absF_vs_MAE_na}
\end{figure}

Figure \ref{fig:scatter_U_and_absF_vs_MAE_na} shows the AE of models trained  at 300$\,\mathrm{K}$ (2a,2c) and at 2000$\,\mathrm{K}$ (2b,2d) as a function of the UQ$_d$ signal (2a,2b) and as a function of the predicted force magnitude (2c,2d).

Observing Figure \ref{fig:scatter_U_and_absF_vs_MAE_na} we note that the AE correlates with both the predicted force magnitude and the direct UQ. Another observation is that the force magnitudes are higher at the higher temperature, as can be expected. This indicates that even though it may seem that the $\vec{UQ_d}_{(i,a)}$ signal is able to produce a prediction for the AE, it might be biased by the force magnitude which by itself is affected by the temperature. When comparing models trained on data with different typical distributions of forces (controlled by temperature), this bias will cause over-confidence when predicting weaker forces, and under-confidence when predicting higher forces. The difference in MAE between train  temperatures was also observed by our earlier works~\cite{kuritz2018size,sivanniv2020}. 

For the Na300$\,\mathrm{K}$ model, we observed an increase in the $UQ_d$ signal when the test dataset temperature is higher from the training temperature (Figure 2a) and we could easily identify a threshold that produces a good separation between in-distribution and OOD model predictions.
However, when we examined the same $\vec{UQ_d}_{(i,a)}$ signal for the Na2000$\,\mathrm{K}$ model and tested on datasets generated with lower temperatures such as 300$\,\mathrm{K}$ in Figure 2b, we observed that the UQ$_d$ signal was lower compared to the 2000$\,\mathrm{K}$ data, despite the model performing worse compared to the Na300$\,\mathrm{K}$ trained model. While the test cases which are shown in Figure \ref{fig:scatter_U_and_absF_vs_MAE_na} through Figure \ref{fig:AL_relative_uncertainty_with_th} are 2000$\,\mathrm{K}$ and 300$\,\mathrm{K}$, we show in the SI (Section S2) the performance of both models when tested on data from a wider range of temperatures, revealing the same pattern displayed here. 

To summarize this part, we showed that the $UQ_d$ signal calculation (and the AE) was higher at higher temperatures regardless of the training temperature of the model (or if it's OOD or not). Therefore this signal could not help us to  separate the in-distribution temperature from the OOD temperature.

\subsection{Relative UQ signal temperature effects}

\begin{figure}[H]
\centering
\includegraphics[width=1.0\textwidth]{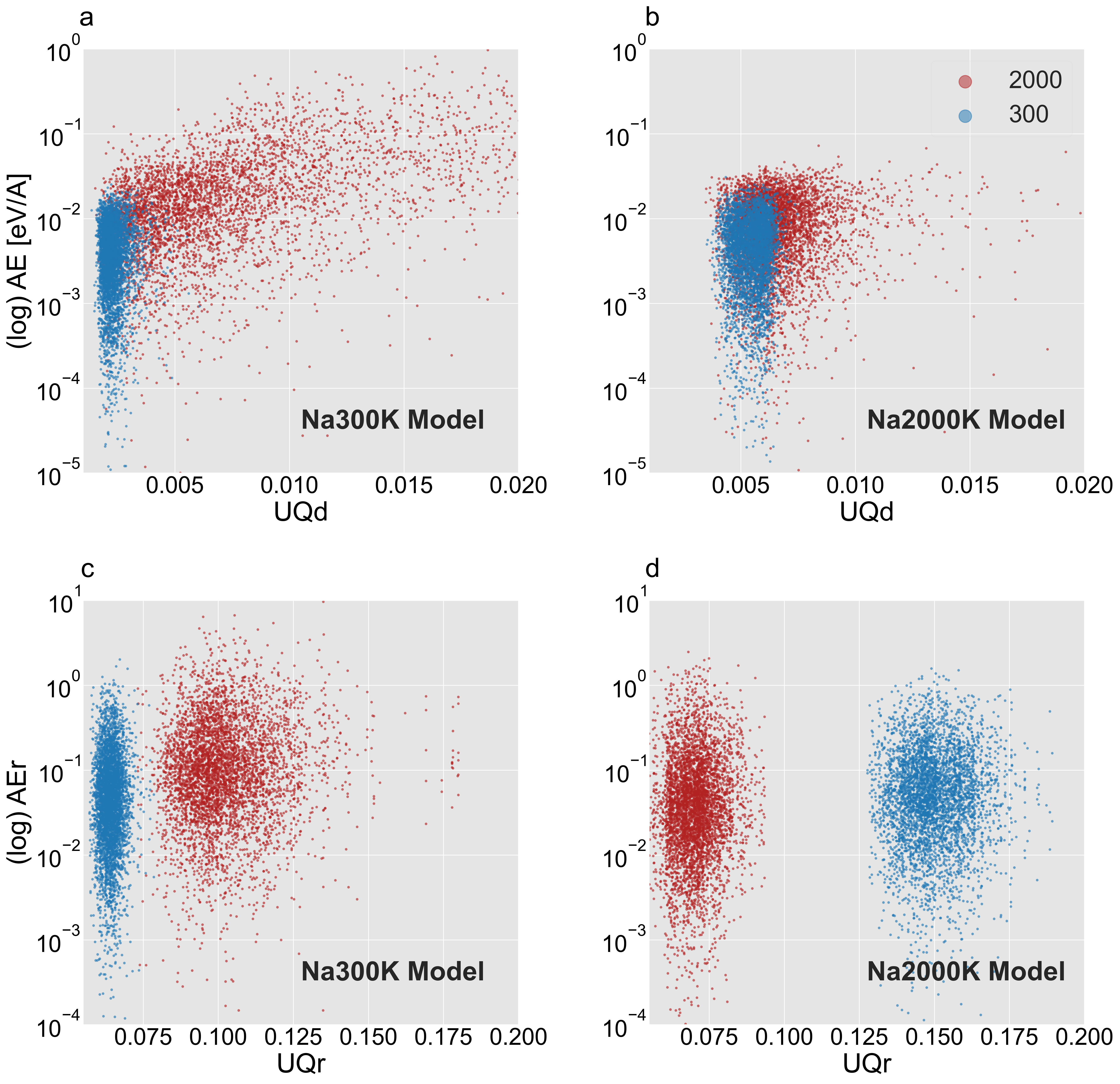}
\caption{The ${UQ_d}$ and the ${UQ_r}$ behavior of models trained and tested on 2 different temperature datasets, Na300$\,\mathrm{K}$(blue) and Na2000$\,\mathrm{K}$ (red), is shown. To improve visual separation, the $UQ_{r}$ signal was normalized as in Eq.~\ref{eq:F_norm_mb8}, calculated using $W_{n}$ of 5, and both the $UQ_{r}$ and the $UQ_{d}$ signals were smoothed as in Eq.~\ref{eq:U_smoothed_calbirated_mb9} calculated using $W_{u}$ of 5. The top sub-figures pair (a,b) show the ${UQ_d}$ signal (Eq.~\ref{eq:UQd_signal_mb4}), against the Absolute Error (AE of Eq.~\ref{eq:AE_defintion_mb2}) on a logarithmic scale. In the second sub figures pair (c,d), we show the ${UQ_d}$ signal against the Relative AE, (${AE_r}$) (Eq.~\ref{eq:rel_AE_mb5}). }
\label{fig:figure_scatter_na_mb3}
\end{figure}

To be resilient to the heteroscedastic behavior of the UQ signal, Gal et al.~\cite{gal2016dropout} and Lakshminarayanan et al.~\cite{lakshminarayanan2017simple} proposed estimating the standard deviation of the prediction as a function of the input, and integrating it to a Gaussian \footnote{Gaussian loss is an example and solutions should be valid for other parametric distributions } distribution loss. A calibration for the estimation of the standard deviation was suggested by Hirschfeld et al.~\cite{hirschfeld2020uncertainty}, fitting a parametric equation to a Gaussian loss. 

In this work, we check whether it is possible to build a good uncertainty estimation to an existing NN model without the addition of a standard deviation head to the model itself and without  assumptions of distribution family on the data.

As discussed in the previous paragraph, both the UQ$_d$ signal and the AE seem to depend on the force magnitude, which depends on the training data temperature. 
Hence, it makes sense to scale the UQ signal by the average force magnitude as described by eq. \ref{eq:U_calbirated_mb7}.
We further smoothed this new relative UQ signal, denoted as ${UQ_r}$, by averaging the force magnitude as described in Eqs. \ref{eq:F_norm_mb8a} through  \ref{eq:U_smoothed_calbirated_mb9}.

Figure \ref{fig:figure_scatter_na_mb3} shows the AE as a function of the $UQ_d$ and ${UQ_r}$ signals. It is evident from the figure that the OOD data, whether it is 300$\,\mathrm{K}$ or 2000$\,\mathrm{K}$, gives on average a higher $\vec{UQ_r}_{(i,a)}$ signal relative to the training data, furthermore, both the OOD and the training data are better clustered by the $UQ_r$ signal. This is in contrast to the $UQ_d$ signal shown in Figure \ref{fig:figure_scatter_na_mb3}(a,b). Hence, the scaling improved the ability of the UQ signal to detect OOD. Another observation from Figure \ref{fig:figure_scatter_na_mb3} is that the UQ$_r$ signal, when considered point-wise, is not necessarily a good predictor of the AE or relative AE.

\begin{figure}[H]
    \centering
    \includegraphics[width=1.00\textwidth]{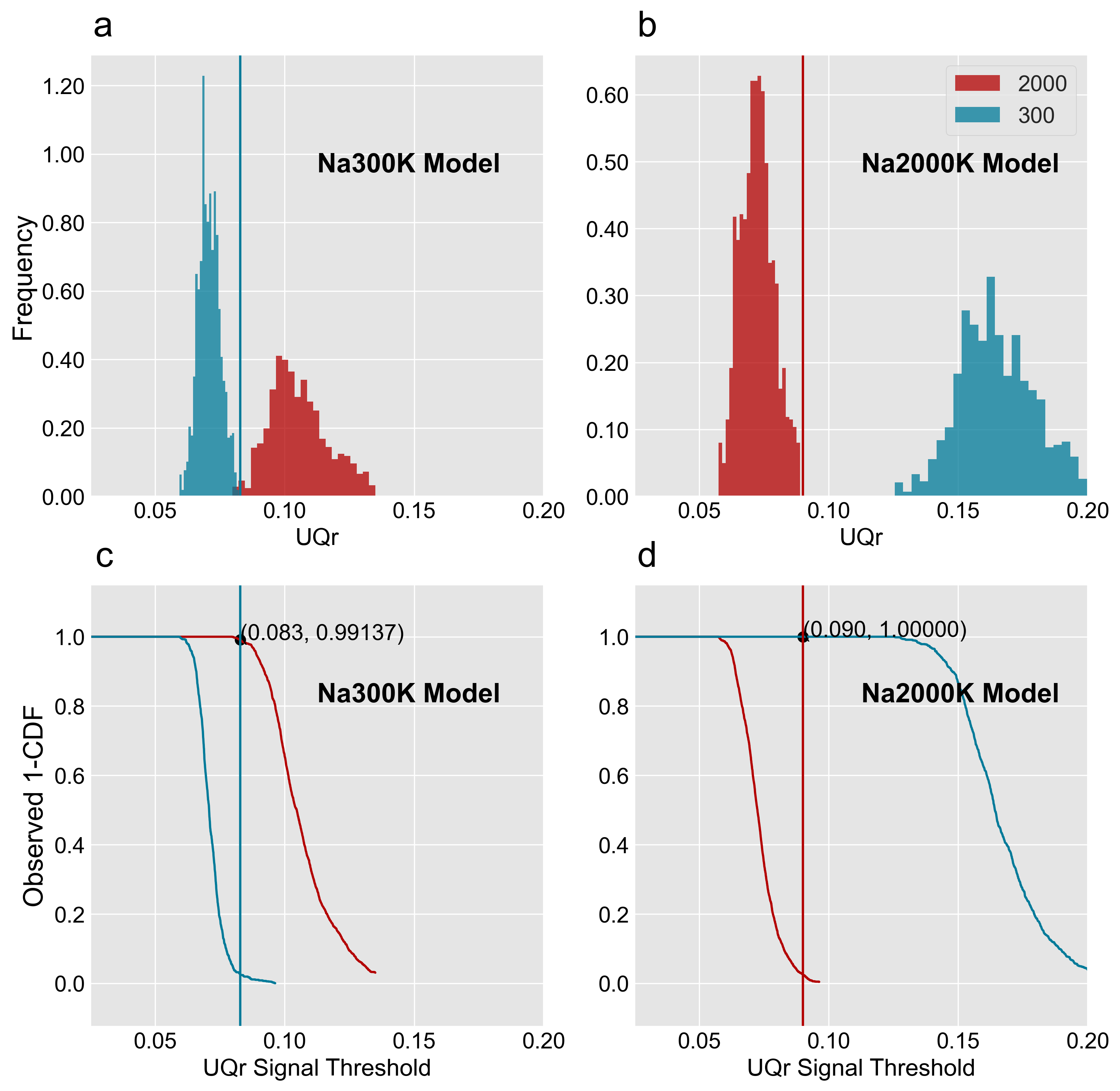}
    \caption{In (a, b) - Distribution of the ${UQ_r}$ (Eq.\ref{eq:U_smoothed_calbirated_mb9}) for Sodium for 300$\,\mathrm{K}$ (a) and the 2000$\,\mathrm{K}$(b) trained models, tested on 2000$\,\mathrm{K}$(red) and 300$\,\mathrm{K}$(blue) data, in bins of 0.01.  In (c,d) - The observed cumulative density function (CDF) obtained from Eq.\ref{eq:P_ood_mb12} measuring the frequency of ${UQ_r}$ signal samples that are above a threshold set at that point. The vertical line is set on the 95th percentile of the relative uncertainty on train temperature for Na300$\,\mathrm{K}$ (a,c) and Na2000$\,\mathrm{K}$ (b,d). }
    \label{fig:na_relative_uncertainty_with_th}
\end{figure}

To use this  $\vec{UQ_r}_{(i,a)}$ signal for decisions, a threshold for OOD detection should be fixed. Ideally, we would like the same threshold for both the 300$\,\mathrm{K}$ data and the 2000$\,\mathrm{K}$ data (and other similar datasets). A possible way to determine such a threshold is to fix a threshold that yields a low probablity of OOD, for data similar to the training data; here we selected a probablity of $0.025(2.5\%)$. Using this method we found a threshold of 0.083 for the 300$\,\mathrm{K}$ data and 0.09 for the 2000$\,\mathrm{K}$ data. Choosing a common threshold of 0.087, we get a high acceptance rate for the temperature that the model was trained with. We now check what happens with the OOD data, For the model that was trained at 2000$\,\mathrm{K}$, all (100\%) the 300$\,\mathrm{K}$ data was above the threshold. For the model that was trained at 300$\,\mathrm{K}$ 99\% of the 2000$\,\mathrm{K}$ data was above the threshold.

 \subsection{Aluminum Bulk and Slab}

\begin{figure}[H]
    \centering
    \includegraphics[width=1.00\textwidth]{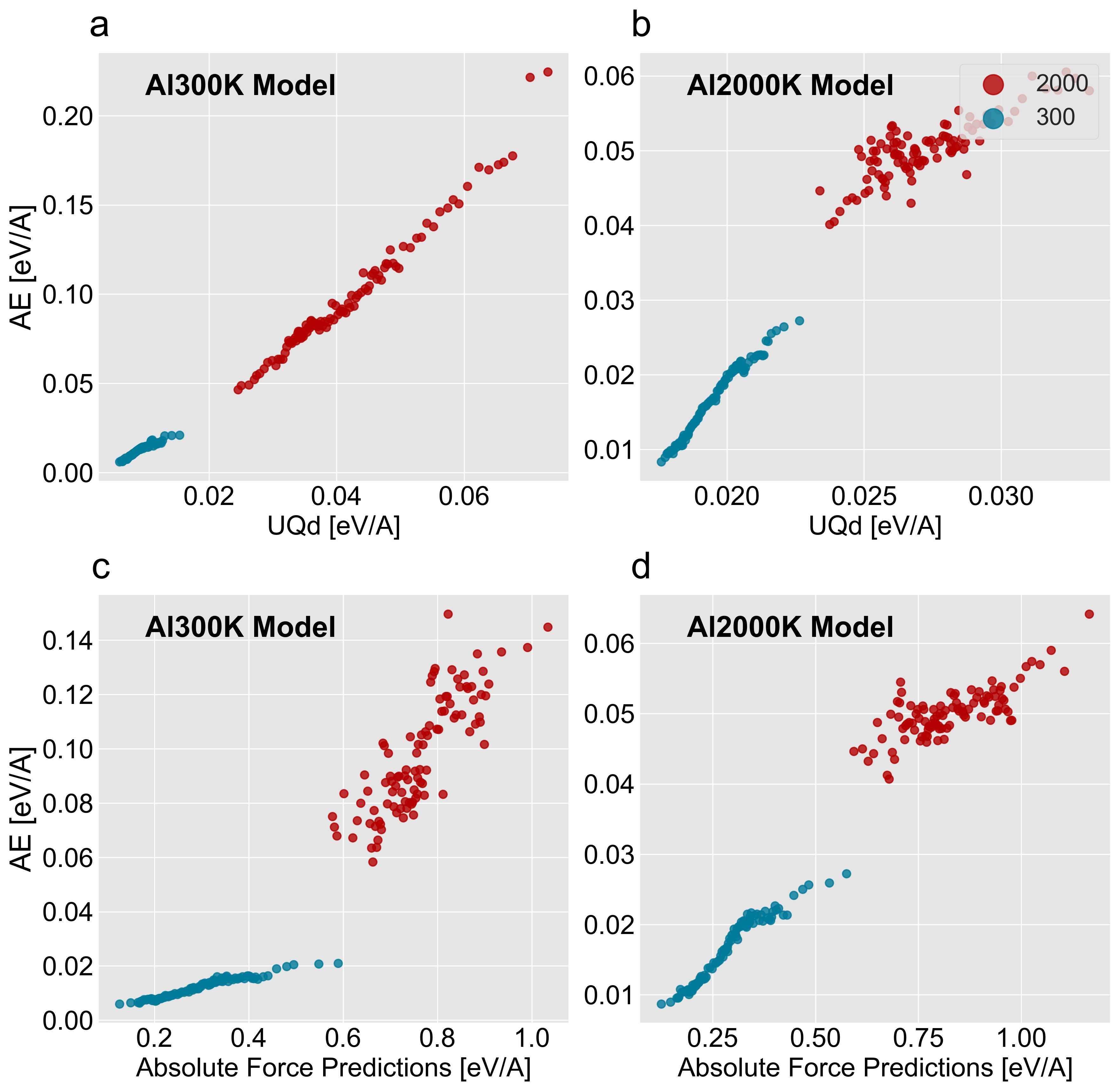}
    \caption{On all graphs, exactly like in Figure \ref{fig:scatter_U_and_absF_vs_MAE_na} only for Aluminum instead of Sodium, the data is grouped into 100 bins according to the x-axis, and the AE is averaged in each bin. (a, b) - Scatter of the AE (Eq.\ref{eq:AE_defintion_mb2}) vs. the direct UQ signal (Eq.\ref{eq:UQd_signal_mb4}) produced for the Al300$\,\mathrm{K}$ (a) and the Al2000$\,\mathrm{K}$ (b) Aluminum trained models, tested on Al2000$\,\mathrm{K}$(red) and Al300$\,\mathrm{K}$(blue) data.  (c,d) a scatter plot of the AE (Eq.\ref{eq:AE_defintion_mb2}) vs the predicted force magnitude (Eq.\ref{eq:F_norm_mb8}) is shown for models trained on Al300$\,\mathrm{K}$ (c) and Al2000$\,\mathrm{K}$ (d). Red is the result of a Al2000$\,\mathrm{K}$ test set and Blue is Al300$\,\mathrm{K}$ test set. }\label{fig:scatter_U_and_absF_vs_MAE_al}
    \label{fig:al_scatter_U_and_absF_vs_AE}
\end{figure}

\begin{figure}[H]
    \centering
    \includegraphics[width=1.00\textwidth]{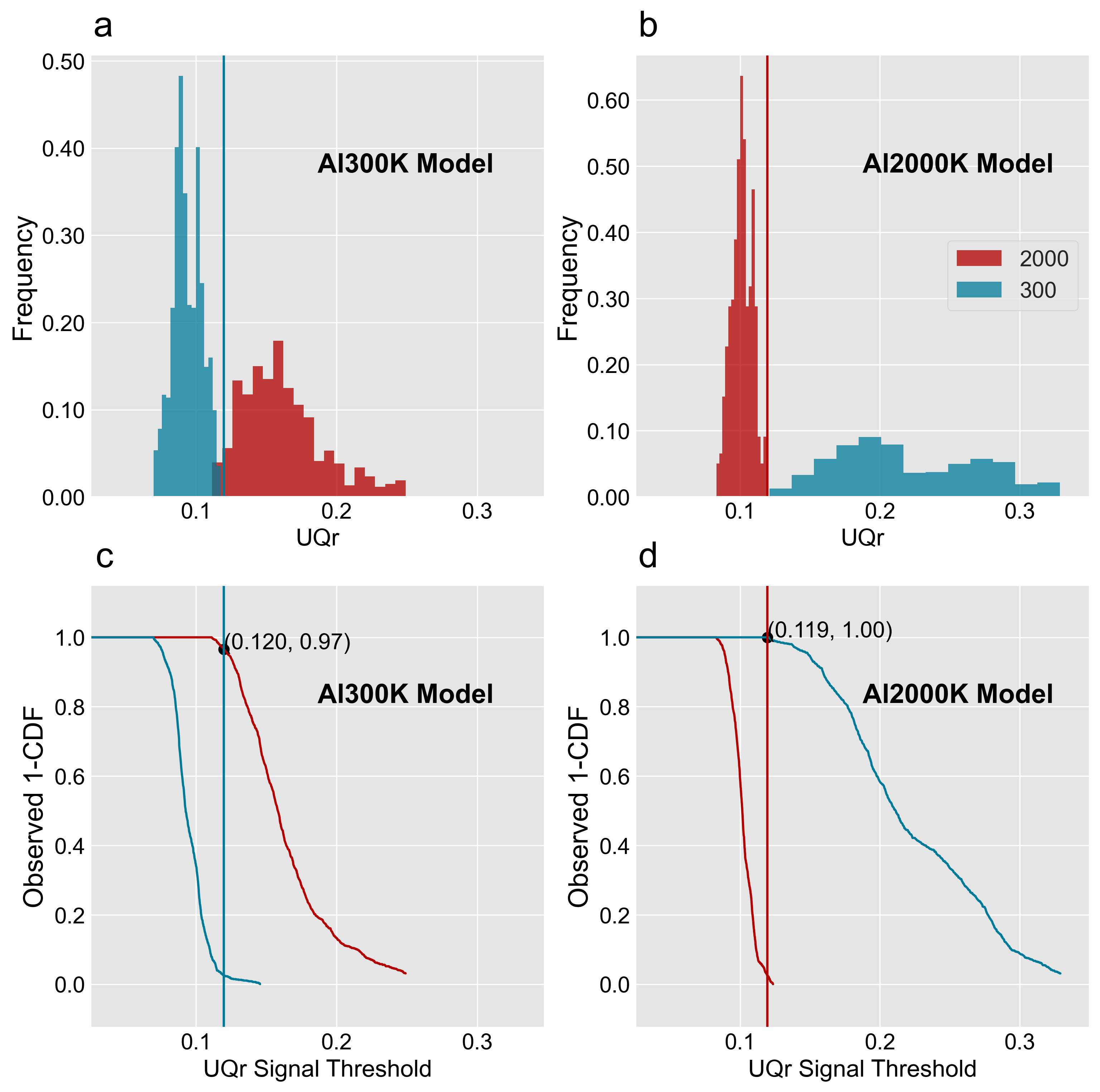}
    \caption{In (a, b) - Distribution of the ${UQ_r}$ (Eq.\ref{eq:U_smoothed_calbirated_mb9}) for Aluminum for 300$\,\mathrm{K}$ (a) and the 2000$\,\mathrm{K}$(b) trained models, tested on 2000$\,\mathrm{K}$(red) and 300$\,\mathrm{K}$(blue) data, in bins of 0.01.  In (c,d) - The observed cumulative density function (CDF) obtained from Eq.\ref{eq:P_ood_mb12} measuring the frequency of ${UQ_r}$ signal samples that are above a threshold set at that point. The vertical line is set on the 95th percentile of the relative uncertainty on train temperature for Al300$\,\mathrm{K}$ (a,c) and Al2000$\,\mathrm{K}$ (b,d). }
    \label{fig:AL_relative_uncertainty_with_th}
\end{figure}

To further validate our approach, we conducted a similar analysis for bulk aluminum (Al), using Al300\,K and Al2000\,K for training data. Figure~\ref{fig:al_scatter_U_and_absF_vs_AE} shows the behavior of both models on $300\,\mathrm{K}$ and 2000$\,\mathrm{K}$ data.

Although the absolute error (AE) values for Al are generally higher than those of Na, their correlation with the force magnitudes follows the same trend. This again indicates that relative uncertainty estimates, rather than direct ones, are more appropriate for this evaluation.

Our observations confirm that the uncertainty signal ${UQ_d}$ retains its dependence on the predicted force magnitudes, as seen previously with Na. More importantly, the UQr signal remains a robust indicator for distinguishing between in-distribution and OOD configurations. We also note that the UQr values for the aluminum models were slightly higher than those of Na. Despite these elevated UQr levels, the computed threshold from Eq.~\ref{eq:P_ood_mb12} enabled effective rejection of OOD configurations. Like the case of sodium, we tried now to put a threshold that will accept the in-distribution data and reject the OOD. As shown in Figure ~\ref{fig:al_scatter_U_and_absF_vs_AE}, similarly to the Na results, for the Al2000$\,\mathrm{K}$ we found a threshold that rejects only 2.5\% of in-distribution and 100\% of OOD, and we found a threshold for the AL300$\,\mathrm{K}$ that rejects 2.5\% of in-distribution and 97\% of OOD. 
For both the Al300$\,\mathrm{K}$ and Al2000$\,\mathrm{K}$ the ${UQ}_{r}$ threshold was around 0.12, which is different than the Na (0.083-0.09) threshold but not very far. Another observation we found is that the $UQ_r$ signal is sensitive to model convergence, it is therefore critical to perform the analysis on well converged models.

\begin{table}[htbp]
	\centering
	\caption{Performance metrics for Na and Al materials at different training and testing temperatures. The table compares prediction accuracy using multiple evaluation metrics (MAE, ${MAE_r}$, $UQ_d$, $UQ_r$) and shows the true labels alongside the model predicted values.}
	\begin{tabular}{|c|c|c|rrrrrr|}
		\hline
		\multirow{2}{*}{X} & \multirow{2}{*}{Train$\,\mathrm{K}$ } & \multirow{2}{*}{Test$\,\mathrm{K}$ } & \multicolumn{6}{c|}{Metrics} \\
		\cline{4-9}
		&  &  & MAE & ${MAE_r}$ & ${UQ_d}$ & ${UQ_r}$ & ${\| \vec{F}^{DFT} \|}$ & ${\| \vec{F}^{\mathrm{Pred}} \|}$ \\
		\hline
		\multirow{4}{*}{Na}
		& 300  & 300  & 0.005 & 0.279 & 0.002 & 0.071 & 0.099 & 0.099 \\
		& 300  & 2000 & 0.045 & 0.415 & 0.009 & 0.107 & 0.288 & 0.251 \\
		& 2000 & 2000 & 0.011 & 0.248 & 0.007 & 0.073 & 0.288 & 0.287 \\
		& 2000 & 300  & 0.006 & 0.323 & 0.005 & 0.167 & 0.099 & 0.101 \\
		\hline
		\multirow{4}{*}{Al}
		& 300  & 300  & 0.012 & 0.287 & 0.009 & 0.095 & 0.289 & 0.289 \\
		& 300  & 2000 & 0.097 & 0.419 & 0.041 & 0.164 & 0.824 & 0.761 \\
		& 2000 & 2000 & 0.050 & 0.365 & 0.027 & 0.102 & 0.824 & 0.821 \\
		& 2000 & 300  & 0.017 & 0.336 & 0.019 & 0.224 & 0.289 & 0.282 \\
		\hline
	\end{tabular}
	\label{tab:NA_and_Al_Bulk}
\end{table}

We next extended our tests to the slab configurations which are described in the dataset part. In order to understand both the error and the UQ locally, we now evaluate both per atom and not per cell, hence most of the averaging and smoothing is performed over data samples for the same atom and direction. 

\begin{figure}[H]
\centering
\includegraphics[width=1.0\textwidth]{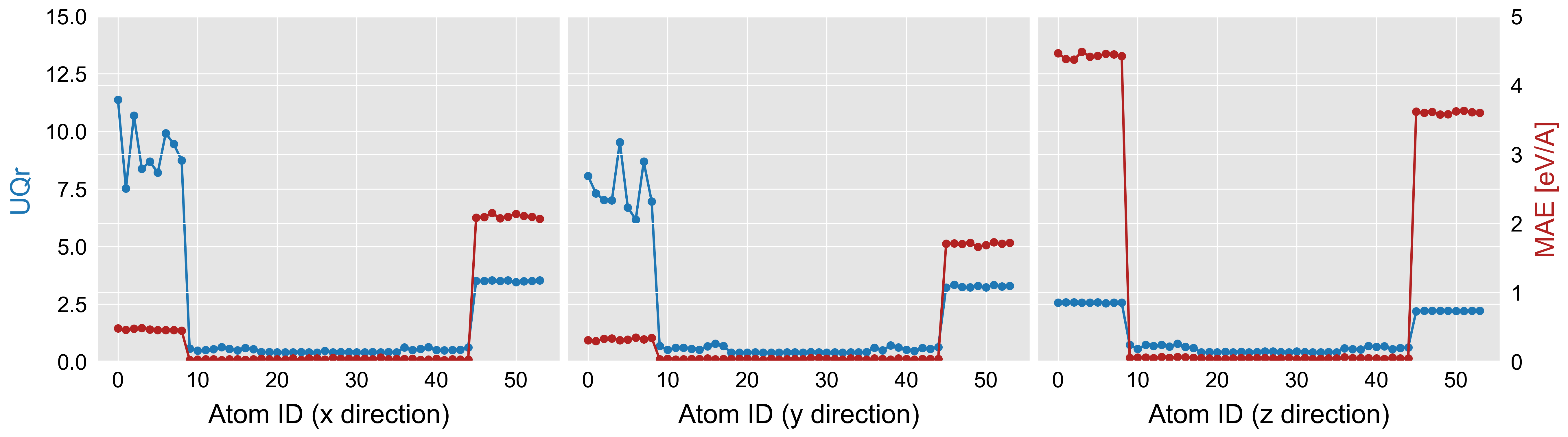}
\caption{\label{fig:Al_slab} The local $UQ_r$ of an Aluminum model trained on 300$\,\mathrm{K}$ bulk configuration, and tested on 300$\,\mathrm{K}$ slab configuration (right axis) and corresponding MAE (left axis). The local $UQ_r$ is calculated using Eq.\ref{eq:U_smoothed_calbirated_mb9} without smoothing over the direction or over the atom in the cell, and smoothed only over samples in time, to isolate the directional components. The local MAE here is not calculated over the whole cell but per atom (doing the average over all data samples). We show in Figure S7 and Figure S8 in the SI the same results for the $MAE_r$ .}
\end{figure}

We used the Al300$\,\mathrm{K}$ bulk trained model to predict the forces and uncertainty on the samples of the Aluminum 300$\,\mathrm{K}$ slab data. 
We show in Figure \ref{fig:Al_slab} the $UQ_r$  of the force prediction in all directions, with $z$ being the direction perpendicular to the slab. The atoms on the top and bottom layers of the slab have a significantly different neighbor environment when compared to the bulk. We therefore, expect both the forces error in that direction and the uncertainty to be high. 
We observed that the ${UQ_r}$increased significantly for atoms located on both surfaces of the slab.
This behavior highlights the model's sensitivity to structural changes, particularly at the surface boundaries. The large MAE which is found for the surface atoms can be attributed to the fact that such a neighbor configuration is not found in the bulk dataset, the error in such cases can become even larger due to the unbounded nature of the ReLU activation function that was used.

{ 

Our findings for the behavior of the AE and UQ for surface atoms agree well with the results of Carrete et al.~\cite{carrete2023deep}. They showed that both the error and the UQ increase significantly for atoms near the free surface of $SrTiO_3$ and decrease rapidly towards the bulk.

}

It is to be noted that for the slab configuration, we have chosen a different set of smoothing configurations \ref{eq:U_smoothed_calbirated_mb9} (atom and direction independent ) which generated the UQ signal with respect to directions. This allows the detection of directional bias between the Cartesian coordinates of the forces.   Specifically, we used $W_{u}$ of 15 and $W_{n}$ of 15 samples with single atom normalizing (the denominator) and smoothing. We increased the smoothing over sample space because we are now using purely local data and would like to differentiated between atoms and directions, therefore refrain from smoothing over the cell atoms by increasing the temporal samples window.

\subsection{Model Training and Number of Samples}

A natural question is how many samples are needed to train a model on a given dataset and what is the relation between the MAE and the ${UQ_r}$. 

As before, we used models that were trained on Na300$\,\mathrm{K}$ data and on Na2000$\,\mathrm{K}$ data, and analyzed the behavior of both the MAE and the $UQ_r$ signal during training as a function of the number of samples used. This was implemented by selecting only $N$ samples of the training data, for example, using only $N=50$ random samples instead of 1000 and repeating the training for several $N$ values. 

We trained around 500 ensembles of 25 experts each with a different $N$ value, and used the same bootstraping approach that was used before. Note that $N$ is not evenly distributed along the x axis and is selected with denser values of $N$ around the inflection points of the graphs. Note that the peaks on the points 595 and 685 are the results of a non-converged expert out of the 25, and we kept it in the dataset for transparency.

\begin{figure}[H]
\centering
\includegraphics[width=1.0\textwidth]{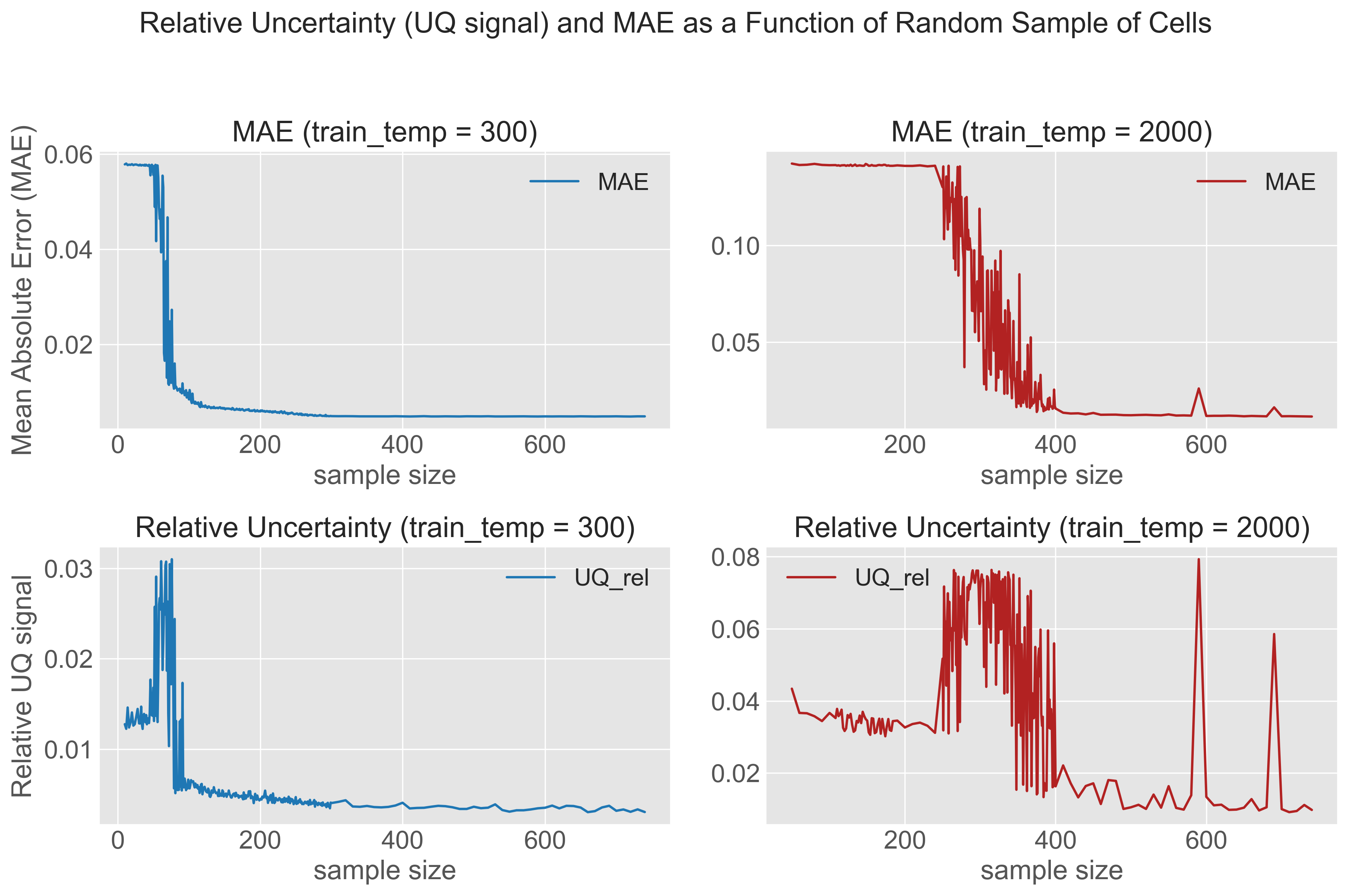}
\caption{\label{fig:figure_learning_curve} MAE (Eq.\ref{eq:mae_pb3}) and ${UQ_r}$ (Eq.\ref{eq:U_smoothed_calbirated_mb9})
 with the sample configuration as in Figure \ref{fig:na_relative_uncertainty_with_th} as a function of the number of training samples for a model trained on Sodium Na300$\,\mathrm{K}$ (left axis) and Na2000$\,\mathrm{K}$ (right axis) datasets.
 }
\end{figure}

 Figure \ref{fig:figure_learning_curve} shows the results for both data sets. For better visualization, we used a higher resolution sampling in areas with of a rapid change in the MAE or ${UQ_r}$ signal. 

It is evident that for both the 300$\,\mathrm{K}$ and 2000$\,\mathrm{K}$ data there is a critical number of samples for the MAE to drop. 
For the 300$\,\mathrm{K}$ data this is happening between 50 to 100 samples, while for the 2000$\,\mathrm{K}$ data it happens between 200 and 400 samples of data.
{In this analysis, the number of training epochs was kept constant in all experiments, regardless of the number of data samples used. This approach makes it easy to isolate the effects that are due to data size changes. Increasing the epochs number would generally allow for a smaller number of training data samples up to some limit.}

Examining the behavior of the ${UQ_r}$ signal we observe an interesting pattern, both in the 300$\,\mathrm{K}$ data and 2000$\,\mathrm{K}$ data there is a jump in the ${UQ_r}$ signal at the point where the MAE starts to go down, this jump is followed by a reduction in the ${UQ_r}$ signal after the MAE converges to the smaller value. A possible explanation for the jump is that at the region where the MAE goes down, not all the experts in the ensemble reached convergence to the smaller MAE and therefore, the ${UQ_r}$ signal, which is based on the standard deviation of the experts, is high. Once the MAE is stabilized all the experts are converged to the lower MAE value and the ${UQ_r}$ signal is also reduced.

We thus suggest that the ${UQ_r}$ signal may be also used to track the transition between the model training operation points. This tracking can be performed using either the direct uncertainty ($UQ_d$) or the ${UQ_r}$ signal. Moreover, in model ensembles with a suddenly steeped learning curve, the ${UQ_r}$ signal should reach a peak, as it represents the standard deviation between 2 operation points. We hence suggest that this behavior, of a "differential" like peak of a standard deviation based uncertainty signal, is a universal property of ensemble methods for uncertainty estimation. This is because regardless of the problem, at a turning point in the error, some of the experts are going to improve their fit level faster than the others.

\section{Conclusions}

In this work, we examined a "black-box" ensemble-based approach for uncertainty quantification (UQ) in atomistic neural networks that predict the forces directly. In this approach, we built an ensemble of identical models that are trained on differently sampled data, hence the model itself can be treated as a "black-box" multi-headed regression model.
We showed that both the MAE and the direct UQ signal were correlated with the force magnitude value, obscuring the ability to detect OOD data. We therefore introduced a ${UQ_r}$ signal, which was scaled by the magnitude of the predicted forces. Although the ${UQ_r}$ signal was not a satisfactory predictor of $MAE$ or ${MAE_r}$, it showed very good performance in detecting OOD. With this ${UQ_r}$ signal we were able to show the detection of OOD for both bulk sodium and aluminum, detecting data that were derived from a different temperature.{ We further demonstrated that the purely spatially local ${UQ_r}$ signal as defined by Eq. \ref{eq:U_smoothed_calbirated_mb9} is useful for the detection of the different environment of the surface layers atoms in an aluminum slab with a model that was trained only on bulk. The use of spatially local UQ signal can be important for surface problems, as demonstrated by Carrete et al.\cite{carrete2023deep}, and also for surface reactions and surface catalysis. This is because there might be a large change in the local atomic environment during the surface chemical reaction. An additional observation for both the direct and relative UQ signals is that they can be noisy when calculated point wise. Therefore, the choice of a data samples smoothing window (as described in Eq. \ref{eq:U_smoothed_calbirated_mb9}) can be important.

Another key issue is the number of data samples which are required for a model to achieve a sufficiently low MAE and ${UQ_r}$ signal. Here, we demonstrated that at lower temperatures the model needed less data samples to converge. In addition, we demonstrated that the $UQ_r$ signal initially rises and then declines around the location where the MAE goes down. We attribute this behavior to the fact that near a turning point in the MAE the pace of the experts convergence to the lower MAE is not uniform. Hence, it is expected that any measure which is based on experts disagreement should jump at this stage of the learning in any problem with a fast enough change of the MAE (or any other measure for the model error) with the number of data points. This suggests the $UQ_r$ signal can also serve as a general proxy for the model's goodness of fit. 

Since the use of ML methods in quantum chemistry is growing rapidly the importance of a reliable UQ signal and the knowledge of its behavior are becoming critical. A good UQ signal should be able to both give a good prediction for the error ( or relative error ) and for the question of whether the simulation data is close to the training data set or OOD. With our ${UQ_r}$ signal we achieved OOD detection but not a good correlation to the actual error. 

While it is clear that models which include error estimation as part of the model can be advantageous, a black box approach has the advantage that it can be used regardless of the model. Hence giving the possibility to both estimate and improve model performance without access to the model itself. This can be applied to a broader set of problems both in computational chemistry and also in completely different fields.

\section*{Acknowledgments}
AN acknowledges funding from the PAZY foundation grant 233/20.

\providecommand{\latin}[1]{#1}
\makeatletter
\providecommand{\doi}
{\begingroup\let\do\@makeother\dospecials
	\catcode`\{=1 \catcode`\}=2 \doi@aux}
\providecommand{\doi@aux}[1]{\endgroup\texttt{#1}}
\makeatother
\providecommand*\mcitethebibliography{\thebibliography}
\csname @ifundefined\endcsname{endmcitethebibliography}
{\let\endmcitethebibliography\endthebibliography}{}

\end{document}